# Nitrogen doping of metallic single-walled carbon nanotubes: *n*-type conduction and dipole scattering

V. Krstić[1,2]*, G.L.J.A. Rikken[1], P. Bernier[3], S. Roth[4], M. Glerup[2,3,5]*

[1]*Laboratoire National des Champs Magnétiques Pulsés, CNRS / INSA / UPS, 143, Avenue de Rangueil, 31432 Toulouse, F*

[2]*Grenoble High Magnetic Field Laboratory, CNRS, 25, Avenue des Martyrs, 38042 Grenoble, F*

[3]*GDPC (UMR5581), Université Montpellier II, Pl. E. Bataillon, 34095 Montpellier, F*

[4]*Max-Planck-Institut für Festkörperforschung, Heisenbergstr. 1, 70569 Stuttgart, D*

[5]*Department of Chemistry, University of Oslo, P.O. Box 1033 Blindern, 0315 Oslo, N*

*These authors contributed equally to this work. Correspondence to krstic@lncmp.org  and marianne.glerup@kjemi.uio.no*

**January 2006**

## Abstract

**The charge transport properties of individual, metallic nitrogen doped, single-walled carbon nanotubes are investigated.  It is demonstrated that *n*-type conduction can be achieved by nitrogen doping.  Evidence was obtained by appealing to electric-field effect measurements at ambient condition.  The observed temperature dependencies of the zero-bias conductance indicate a disordered electron system with electric-dipole scattering, caused mainly by the pyridine-type nitrogen atoms in the honeycomb lattice.  These results illustrate the possibility of creating all-metallic molecular devices, in which the charge carrier type can be controlled.**

In the past, the electronic properties of multi- and single-walled carbon nanotubes (MWNTs and SWNTs) have been intensively investigated. The primary reason is that they represent an almost perfect model for fundamental research due to their unique one-dimensional (1D) electronic structure. For the same reason nanotubes are promising candidates for applications in molecular devices (*1, 2*), provided that the electronic properties, and in particular the type of charge carriers, can be controlled. Major efforts have been undertaken to this end ranging from chemical modifications,(*3*) exposure to different gaseous atmospheres (*4-7*) and immersion in electrolytes (*8*). Another possibility for tailoring the electronic system of the nanotubes is their deliberate doping. In this case, doping refers to the substitution of a carbon atom with other elements such as boron or nitrogen, and is expected to have significant impact on the charge transport properties of the nanotubes.(*9*) The successful doping of multi-walled nanotubes with nitrogen has already been achieved (*10-13*), and confirmed by thermopower measurements (*12-14*). It was recently shown that it is possible to grow directly nitrogen-doped SWNTs i.e. with no post-processing (*15*). Among these, the metallic SWNTs are of particular interest for the development of all-metal based molecular nano-electronics, since they represent excellent building blocks for devices with low power consumption in combination with large current densities. However, in standard devices from SWNTs *p*-type conduction is always observed. This results in severe limits on possible device architectures due to the absence of *n*-type conducting SWNTs.

Here we present for the first time investigations on individual metallic nitrogen doped SWNTs, showing that intrinsic *n*-type conduction can be achieved. Furthermore, in the low-energy limit, the charge carriers of nitrogen doped SWNTs are found to display a temperature dependent scattering mechanism, which is attributed to electric-dipole moment interactions. These two experimental observations are correlated to the different bonding configurations of nitrogen in the carbon-host lattice. The nitrogen-doped SWNT raw material used for these experiments was synthesized by using a recently developed arc-discharge method yielding nanotubes with an average tube diameter of ~1.2 nm.(*15*) The average content of nitrogen incorporated into the tubes was around 1 atom%, as determined by electron energy loss spectroscopy. In particular, the results imply that the nitrogen is mainly bonded in two ways, pyridine- and graphite-like, *c.f.* Fig. 1.(*13,15,16*)

FIGURE 1

Individual nanotubes where contacted in a two-terminal configuration by standard electron-beam lithography (electrode material AuPd alloy on top of the SWNT, electrode distance 200 nm).(*17*) In Fig. 2 the transfer characteristic of two different nitrogen-doped SWNT samples, measured at room temperature, are shown (back-gate electrode separated by a 200 nm thick $SiO_2$ layer). As usual for metallic-like nanotubes, the observed variation of the current because of the externally applied electric field was rather weak.

FIGURE 2

Both negative and positive slopes of the gate response are observed, corresponding to *p*- and *n*-type conducting nitrogen-doped SWNTs, respectively. This is remarkable, because the nitrogen atoms are expected to act as electron donors. *P*-type conduction was observed in the majority of the samples investigated. The prevalence of *p*-type conduction in nitrogen doped SWNTs can be rationalized by a compensation of the number of additional, free electrons provided by the nitrogen donor atoms by the number of additional acceptor states (environmental doping) (*18,19*), assuming that the number of acceptor states is larger. However, this simplified picture is misleading for SWNTs: Nitrogen atoms are five-valenced and are incorporated into to the carbon lattice of the nanotubes used in the present work mainly in graphite- and pyridine-like ways.(*15*)  For graphite-like (substitutional) nitrogen all three $sp_2$-orbitals are used to form covalent bonds with the carbon neighbors.(*13*) Thus, the remaining valence electrons occupy the $p_z$-orbital of the nitrogen. Such nitrogen atoms can, in principal, be ionized and the liberated electrons can therefore participate as excess electrons in the delocalised $\pi$-system.  However when considering the time average such electrons from these nitrogen are still to a certain degree localized (and more so in the case of semiconducting tubes).(*20*) In contrast, pyridine-like nitrogen can not be ionized under ambient conditions, because the nitrogen is $sp_2$-hybridised and two of its five valence electrons are in the $sp_2$-orbital, which does not form a $\sigma$-bond to its carbon neighbors.(*13*)  Therefore, this lone-pair will not participate in the conjugated bond system.(*21*) The two types of nitrogen bonding influence the thermodynamics and electrostatics of the SWNTs differently.  At this point it is convenient to recall that the current investigation dealt exclusively with metallic tubes that are zero-gap semiconductors/semi-metals with a vanishing band overlap and a weak electric field-effect.(*22,23*)  In undoped, electrically contacted metallic SWNTs the environmental doping, including workfunction mismatches (*18,19*), modifies the position of the Fermi energy.  From being initially at the charge neutrality point, the nanotube's Fermi level shifts (varying with diameter, helicity and electrode material) into the $\pi$-band, under ambient conditions. In our samples, where we used a AuPd alloy as contact material, the shift is estimated to be ~0.6 eV.(*19,24,25*)  On the other hand, the graphite-like nitrogen, whose excess electrons participate in the delocalised $\pi$-system, leads analogously to a Fermi-level shift towards the $\pi$*-band. The upper limit of this energy shift can be estimated to be smaller than 0.5 eV for our nanotube samples which contain about 1 atom% of nitrogen, according to a recent calculation on SWNTs containing about 2.8 atom% of graphite-type nitrogen (*21*). However, a part of the nitrogen atoms in our samples is of pyridine-like type.

From the electron energy loss spectra of the doped SWNTs (*15*) it is estimated that the amount of graphite-like and pyridine-like nitrogen are comparable. Since the lone-pair of pyridine-like nitrogen does not interact with the delocalised $\pi$-system, it is not contributing to the shift of the Fermi-energy (*21*), setting 0.5 eV to be even a stricter energy border-line. Consequently, considering the magnitudes of the Fermi-level shifts caused by the graphite-like nitrogen and environmental doping, a maximum difference of the order of 100 meV can be estimated in favor of the environmental doping. In other words, only SWNTs with sufficiently high concentration of graphite-like nitrogen can show *n*-type conduction under standard device configuration and ambient conditions. Therefore, mainly *p*-type responses are expected, in agreement with our experimental findings.

Beside the doping effect, the incorporated nitrogen atoms represent potential scattering centers for the conducting charges. In order to investigate their action, the low energy charge transport of *p*-type conducting tubes is investigated, implicitly proving that they are nitrogen-doped and confirming the above considerations. The insets in Fig. 3 show the zero-bias conductance, $G$, versus temperature, $T$, for three samples. In all cases a clear suppression of the conductance towards lower temperature is observed. The main graphs in Fig. 3 show the same plot in a double-logarithmic representation. The straight lines are the theoretical fits for the tube's electron system, assumed to be in a Luttinger-liquid (LL) state: $G(T) \sim T^{\gamma}$, where $\gamma$ is a function of the interaction strength between the charge carriers.(*26*) Sample A (Fig. 3a) exhibits a rather weak electrical coupling to the electrodes (resistance about 108.1 k$\Omega$) and at higher temperature appears to be in good agreement with the LL picture, with an exponent $\gamma \approx 0.77 \pm 0.05$. This value is in contradiction with previous theoretical and experimental values of, $\gamma$, for a weak electrical coupling to the electrodes as in the present case.(*26*) Furthermore, the measurement point close to 4.2 K deviates significantly from the LL fit. For sample B a clear deviation from the LL scenario (straight line) is once more observed.

FIGURE 3a,b,c

In the case of Sample C a power-law relation between $G(T)$ and $T$ seems to be present at higher temperature. Towards lower temperature, the data deviate significantly from this power-law dependence. A detailed analysis of our experimental data, assuming a crossover from a LL to a disordered wire with neutral scatters reveals inconsistencies. (27,28) The linear part of the graph yields an exponent $\gamma \approx 0.2$ giving a value of about 0.4 for the dimensionless interaction strength, $g$, between the charge carriers ($g = N(\gamma\pi)^2/2$; $N$ the number of conducting channels).(27) For the non-linear part a value of $g \approx 3.9$ was obtained from the data, by applying the expression: $G(T) \sim \exp(-\chi/T^{0.5})$, which describes the temperature dependence of the conductance in the disordered wire.(27) Here, $\chi^2 = \varepsilon_c = g/\pi N\tau$ (units $h/2\pi = k_B = 1$) is the characteristic energy accounting for the electrostatic interactions of the charge carriers. The momentum relaxation time $\tau$ is assumed to be constant.(27) The discrepancy in the interaction strengths suggests that the theory used for the data analysis can not account for the experimental results obtained on the nitrogen-doped, single-walled tubes. Indeed, the conclusions drawn from the charge transport results at ambient conditions do support this conclusion. The pyridine-type nitrogen is likely to posses an electric-dipole moment analogous to pyridine, since its lone-pair is not participating to the delocalised electron system. However, because of the carbon host, the electric dipole's magnitude is expected to be significantly smaller than for a free pyridine-molecule (2.215 D).(29) For graphite-like nitrogen the situation is more complex. This type of nitrogen is not bonded in a completely planar way, because of the curvature of the carbon lattice. Furthermore, the excess electron is weakly localized, meaning that a dipole moment could exist. This dipole moment should be much smaller than the dipole moment in a nanotube with pyridine-type dopants. Future detailed theoretical studies should address this particular topic, which is beyond the scope of this work. Consequently, the nitrogen atoms, or at least those of the pyridine-type, can not be regarded as neutral scatterers anymore but as electric-dipole scatterers. Thus, the momentum relaxation time, $\tau$, is temperature dependent, because of the thermal fluctuation of the electric dipole moments.(*30*) These considerations require that the correction has to be small and can be treated to a first approximation by $\tau(T) = \tau_0(1 - \lambda_\tau T)$ with $0 < \lambda_\tau \ll 1$ $K^{-1}$. On that basis and following Ref. 27 and including the above scattering effects, an phenomenological expression for $G(T)$ can be suggested:

$$G(T) \sim (1 - \lambda_\tau T)^{-\gamma_t} \cdot \exp\{-\chi_\tau/(T - \lambda_\tau T^2)^{0.5}\}. \qquad (1)$$

The dashed-dotted lines in Fig. 3 represent fits of the data according to Eq. 1. In table 1 all results on the fitting parameters are summarized. The fit and the experimental data are in reasonable good agreement. Positive values for $\lambda_\tau$ of the order of magnitude of $10^{-3}$ $K^{-1}$ were found, in full agreement with the foregoing considerations. Some data-point fluctuations around the theoretical fits are observed suggesting possible higher order scattering effects, which are not taken into account and should be subject of detailed future theoretical studies on nitrogen-doped SWNTs. However, the fundamental phenomenon that nitrogen atoms can act as electric-dipole scatters in a carbon nanotube lattice, leading to Eq. 1, remains valid. In particular, this data analysis on *p*-type conducting samples confirms that indeed nitrogen atoms are present in all SWNT samples investigated.

TABLE 1

In summary, we have performed electrical-transport measurements on metallic nitrogen-doped single-walled carbon nanotubes under ambient conditions and in the energy limit of low temperatures and zero bias. At room temperature, the results demonstrate, that $n$-type conduction can be achieved, depending on the graphite-type nitrogen concentration and the device configuration (environmental doping). Towards lower temperatures the zero-bias conductance is suppressed. It was demonstrated that the charge transport is better described by a disordered wire with nitrogen atoms acting as electric-dipole scatters than by a Luttinger-liquid state. Pyridine-type nitrogen is suggested to be the main contributor to this scattering process. A phenomenological expression for the zero-bias conductance was proposed, giving a good agreement with the experimental data. In turn, this proves that the tubes are doped with nitrogen, thus demonstrating that electrical transport also is a unique tool for probing low concentration doping in nanotube systems. One can envision that an increased concentration of graphite-like nitrogen relative to pyridine-like nitrogen or the substitution of higher-valence atoms than nitrogen in metallic single-walled carbon nanotube can lead to a reliable method to synthesis $n$-type metallic conductors and can open the field of all-metal molecular electronics.

This work was supported by the European contract no. HPRN-CT-2000-00157.

**Figure captions**

**Fig. 1.** Top: Schematic view of a nitrogen-doped SWNT as used in our measurements. Black and red balls correspond to carbon and nitrogen atoms, respectively. Bottom: The two main ways nitrogen is bonded, graphite- and pyridine-type. Pyridine-type nitrogen is always accompanied with a stabilising defect in the carbon lattice.

**Fig. 2.** $n$- and $p$-type conducting in nitrogen doped nanotubes. Response of the current to the applied electric field at room temperature. The source-drain voltage was set to 250 µV. Curves for negative source-drain bias are equivalent (not shown). All curves exhibit fluctuations (fine-structure), which are also observed in undoped carbon nanotubes.

**Fig. 3.** Zero-bias conductance $G$ vs. temperature $T$ of three $p$-conducting nitrogen doped SWNTs in double-logarithmic presentation. Insets: $G(T)$ with linear axis. **a**, For sample A the data points above 5 K seem to coincide with the dotted line which represents a Luttinger-liquid fit. However, a better agreement is obtained by the dashed-dotted line which is based on the characteristics of a disordered state with $\tau(T) = \tau_0(1 - \lambda_\tau T)$. All points have been corrected for the possible Coulomb-Blockade contribution in the temperature range measured. **b**, Sample B: The dotted line corresponds to a Luttinger-liquid fit whereas the dashed-dotted line is a fit based on Eq. 1 showing a significantly better matching with the experimental data. **c**, Zero-bias conductance of sample C measured down to 2.7 K. The dotted line corresponds to a Luttinger-liquid fit. The dashed-dotted line is a fit based on a disordered state with $\tau(T) = \tau_0(1 - \lambda_\tau T)$ and matches considerably better with the experimental data.

**Table 1.** Charge transport parameters

| sample | $R_{2P}$ [k$\Omega$] | $\gamma_r$ | $\chi_r$ [K$^{0.5}$] | $\lambda_r$ [$10^{-3}$ K$^{-1}$] |
|--------|------------|-----------|------------|------------|
| A | 108.1±0.2 | 0.77±0.05 | 8.97±1.86 | 3.3±0.5 |
| B | 8.5±0.1 | 0.13±0.06 | 1.05±0.04 | 2.9±0.7 |
| C | 30.3±0.1 | 0.20±0.06 | 1.16±0.06 | 6.1±0.4 |

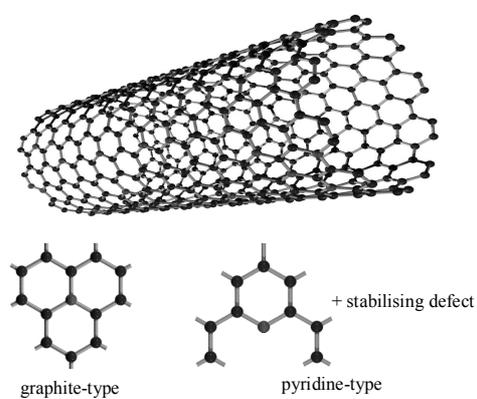

graphite-type        pyridine-type

**Figure 1**

V. Krstić et al.

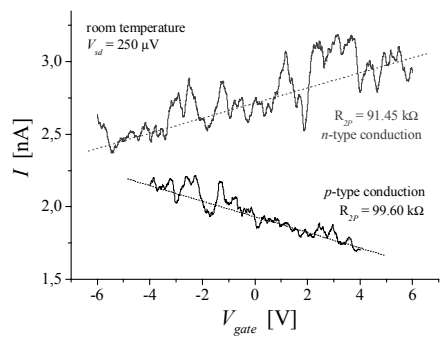

**Figure 2**

V. Krstić et al.

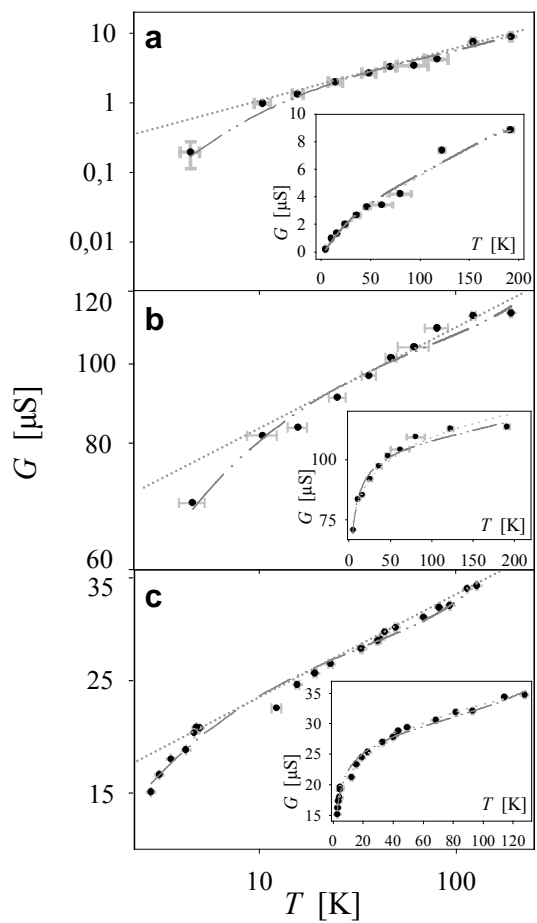

**Figure 3 a, b, c**

V. Krstić et al.